\documentstyle[preprint,epsf]{jpsj}

\def \virg{\;\;,}
\def \point{\;\,.}

\def \e{{\rm e }}
\def \i{{\rm i }}
\def \d{{\rm d}}

\def \V2c{V_{\rm 2c}}

\def\ggs{\buildrel\textstyle > \over {\hbox{\raise0.2ex\hbox{$\sim$}}}}
\def\lls{\buildrel\textstyle < \over {\hbox{\raise0.2ex\hbox{$\sim$}}}}
\def\gsim{\,\lower0.75ex\hbox{$\ggs$}\,}
\def\lsim{\,\lower0.75ex\hbox{$\lls$}\,}

\def \bk{{\bf k}}
\def \bq{{\bf q}}

\def\runtitle{
 Charge fluctuation  induced  superconducting state 
 in two-dimensional quarter-filled electron systems  
 }
\def\runauthor
{}

\title{
 Charge fluctuation  induced  superconducting state 
 in two-dimensional quarter-filled electron systems  
 }

\author{
Akito {\sc Kobayashi}$^{1)}$, 
  Yasuhiro  {\sc Tanaka}
$^{2)}$, 
    Masao  {\sc Ogata}
 $^{2)}$ 
and 
 Yoshikazu  {\sc Suzumura}$^{1)}$
}

\inst{
1)Department of Physics, Nagoya University, Nagoya 464-8602 \\ 
2)Department of Physics, University of Tokyo, Hongo, 
Bunkyo-ku, Tokyo 113-0033 \\
}

\recdate{\hspace{35mm}}

\abst
{
We investigate the superconducting (SC) state  
 in  two-dimensional  extended Hubbard model  at a  quarter-filling
    with the on-site repulsive interaction ($U$) 
     and the nearest-neighbor one ($V$), which lead to 
     the spin fluctuation and the charge fluctuation, respectively.  
 The effect of the charge fluctuation on the onset of the SC state 
  is calculated   by estimating  the pairing interaction    
    within the random phase approximation.  
 Solving the linearized Eliashberg equation 
 with  the anomalous self-energy 
     as the function of both momentum and frequency, 
  we examined the singlet  SC state on the plane of  
    $U$ and $V$ with the fixed temperature. 
 We found a novel result that 
 the SC state  with the  $d_{xy}$ symmetry  retains 
   for most parameters of $U$ and $V$, although 
     the pairing interaction changes the sign 
     from repulsive one  into  attractive one  with increasing $V$. 
The reentrant transition as the function of temperature occurs 
  close to the onset of the SC state.  
}

\kword
{d-wave superconducting state, spin fluctuation,  charge fluctuation,
 nearest-neighbor repulsive interaction, extended Hubbard model
  }

\begin{document}
\newcommand{\vct}[1]{\mbox{\boldmath\(#1\)}}
\sloppy
\maketitle

Organic conductors exhibit  the variety of superconducting (SC) 
states, as found 
   in quasi-one-dimensional (Q1D) materials of Bechgaard salts, 
\cite{Jerome} 
  or quasi-two-dimensional (Q2D) materials of  ET salts.
  \cite{H_Mori,Hotta_JPSJ03} 
 It has been often  asserted that the spin fluctuation  is  
  the origin of the SC state 
   since the  spin density wave state is located 
   next to the SC state  in the phase diagram. 
   However there are some  ET salts  suggesting    
   another possibility that  
   the charge fluctuation could be  the origin of the SC state. 
  In   $\theta$-ET$_2$X salts,    
  where the dihedra angle between two adjacent molecules 
   changes with the choices of the anion X, it has been shown that 
   the state with the charge ordering moves into the SC state 
    with decreasing the dihedral angle
  \cite{H_Mori}  
     corresponding to 
     the increase of transfer energy. 
   It is noteworthy that this SC state is located besides the charge 
    ordering state. 
Moreover there is 
   the recent observation of the SC state in  the Q2D conductor, 
    $\beta^{\prime}$ -ET$_2$ICl$_2$, which  appears 
   under  the pressure of $\sim$ 8 GPa.
 \cite{Taniguchi_JPSJ03}  
 This indicates the increase of the nearest-neighbor repulsive 
  interaction in addition to the increase of the transfer energy.

 In 2D systems, 
 there are several theoretical work for  the SC state,  which is  based on 
 the FLEX ( fluctuation exchange ) approximation.
\cite{FLEX}
Especially, 
 the spin fluctuation as the origin of the singlet SC state 
   has been calculated for several  transfer energies. 
 By  using a model with  
 the on-site repulsive interaction  $U$ and  
   a variation  from  the square lattice 
     to   the triangular lattice,
   \cite{Kontani_PRB03}
 it has been shown 
 that the SC state at half-filling 
  is mainly determined by the state   
   with  the  $d_{x^2-y^2}$ symmetry. 
On the other hand,
 in the presence of the nearest neighbor repulsive interaction $V$, 
  there appears   the SC state with 
the $d_{xy}$ symmetry due to the charge fluctuation,
 as shown by the RPA (random phase approximation)
 \cite{Scalapino} 
  with a choice of $U$ and $V$ 
  and by the slave-boson theory with a quarter-filled band.
  \cite{Merino}
  However, it is not yet clear how the interplay of $U$ and $V$ 
  leads to a crossover from the spin fluctuation to the charge fluctuation 
  especially for the  quarter-filled band.  
   
   In the present paper, such a role of charge fluctuation  is 
    studied by using 
     the two-dimensional extended Hubbard model 
  at the quarter-filling of hole doping. 
      The case of the square lattice 
       is examined as the first step 
      since the transfer integral in  organic conductors 
        is very complicated.  
   Within  RPA, 
    \cite{RPA_1,RPA_2,Scalapino}     
 we examine  a phase diagram 
     of SC,  SDW (spin density wave) and CDW (charge density wave) 
      on  the plane of $U$ and $V$
       and also these states as  the function of temperature.  

We consider the expended Hubbard model  given  by
\begin{eqnarray} 
\label{eq:Hamiltonian}
H = \sum_{{\bf k} \sigma} \varepsilon_{\bf k} c_{{\bf k} \sigma}^+ c_{{\bf k} \sigma} +U \sum_i n_{i\uparrow} n_{i\downarrow} 
 +  V \sum_{<ij>} n_i n_j \virg
\end{eqnarray}
 where quantities  $U$ and $V$ are the coupling constants for 
  on-site and nearest-neighbor repulsive
  interactions.
 The quantity  $\varepsilon_{\bf k}$ denotes
   the band  energy of the 2D square lattice, 
 $\varepsilon_{\bf k}=-2t(\cos k_x a+\cos k_y a) -\mu $, 
 where 
 $t$ is the nearest-neighbor transfer energy  
 and  $\mu$ is determined to obtain the 3/4-filled band, i.e., 
 1/4-filled hole band. 
In eq.~(\ref{eq:Hamiltonian}), 
      $N_{L}$ is the total number of the lattice site 
    and    $a$ is   the lattice constant.  
    We set  $t=1$,  $a=1$ and $k_{\rm B}=1$.

First, we calculate  spin and charge susceptibilities given by 
\begin{eqnarray}
        \label{eq:def_chi}
 & & \chi^{\rm A} (\bq, \omega_l)
   =  \frac{1}{2N_{L}}
   \sum_{\bk, \bk^{\prime},\sigma, \sigma^{\prime}}
   \int_0^{1/T} \d \tau f_{\sigma} f_{\sigma^{\prime}} 
                         \nonumber \\
 & & \left< T_{\tau} c^{\dagger}_{\bk + \bq, \sigma}(\tau) c_{\bk,\sigma}(\tau)
     c^{\dagger}_{\bk^{\prime},\sigma^{\prime}}(0) 
     c_{\bk^{\prime}+\bq, \sigma^{\prime}}(0) 
      \right> 
      \e^{ \i \omega_l \tau}
  \virg 
\end{eqnarray}
 with $T$ and 
 $T_{\tau}$ 
  being the temperature and the time ordering operator of the imaginary time. 
  The spin susceptibility is given 
  by  $A=s$ and 
  $f_{\sigma} =1 (=-1)$ for $\sigma = \uparrow (= \downarrow)$ 
 while  
 the charge susceptibility is given 
  by  $A=c$ and   $f_{\sigma} =1$.   
The irreducible susceptibility $\chi^0 ({\bf q}, \omega_l )$,  
which denotes eq.(\ref{eq:def_chi}) without the interactions, 
is given by 
\begin{equation}
        \label{eq:chi0}
\chi^0 ({\bf q}, \omega_l )
 = -\frac{T}{N_L}\sum_{{\bf k},m}
  G^0 ({\bf k}+{\bf q}, \epsilon_m +\omega_l )
  G^0({\bf k}, \epsilon_m ) \virg 
\end{equation}
where 
 $\omega_l$ $(\epsilon_m )$ is the Matsubara frequency 
for boson (fermion) and 
$
G^0 ({\bf k}, \epsilon_m )
 = [{\rm i} \epsilon_m - \varepsilon_{\bf k} ]^{-1}
$.
Treating $U$ and $V$ terms of eq.~(\ref{eq:Hamiltonian})
 within RPA,
 eq.(\ref{eq:def_chi})
 is calculated as 
\begin{equation}
 \label{eq:chi_s}
\chi^s ({\bf q}, \omega_l ) = 
 \chi^0 ({\bf q}, \omega_l )/(1-U\chi^0 ({\bf q}, \omega_l ))
   \virg 
\end{equation}
\begin{equation}
  \label{eq:chi_c}
 \chi^c ({\bf q}, \omega_l ) = 
 \chi^0 ({\bf q}, \omega_l )
   /(1+(U+2V({\bf q}))\chi^0 ({\bf q}, \omega_l ))
    \virg 
\end{equation}
where $V({\bf q})=2V(\cos q_x a +\cos q_y a)$.

Next,
  we derive the   pairing interaction, 
  $P({\bf q},\omega_l)$, 
  corresponding to the vertex for the singlet state 
\begin{equation}
  \label{eq:def_pairing}
  P({\bf q},\omega_l) 
   \int_{0}^{1/T} \d \tau 
    c^{\dagger}_{\bk - \bq, \uparrow}(\tau) 
     c^{\dagger}_{ - \bk + \bq, \downarrow} (\tau)
     c_{-\bk \downarrow}(0) c_{ \bk\uparrow}(0) 
\e^{\i \omega_l \tau}
    \virg 
\end{equation}
    which is  used in the Eliashberg equation 
     for the SC state. 
   In terms of eqs. (\ref{eq:chi_s}) and  (\ref{eq:chi_c}),  
 the pairing interaction $P({\bf q}, \omega_l )$ 
  is obtained as 
\begin{eqnarray}
    \label{eq:pairing}
P({\bf q}, \omega_l ) 
  & & =  U +V({\bf q}) + \frac{3}{2}U^2 \chi^s ({\bf q}, \omega_l ) 
               \nonumber \\
        & &       -(\frac{1}{2}U^2 +2U V({\bf q})
                 +2V({\bf q})^2) \chi^c ({\bf q}, \omega_l ) \point
\end{eqnarray}

The SC state is determined by the anomalous self-energy, 
$\Sigma^a ({\bf k}, \epsilon_m )$, defined by
\begin{eqnarray}
    \label{eq:self_energy}
   \sum_{{\bf k^{\prime}} \epsilon_{m^\prime}} \int \d \tau 
 \left<T_\tau  c_{\bf k^{\prime} \uparrow }(\tau) 
             c_{\bf -k^{\prime} \downarrow }(0) \right> 
                              \e^{\i \epsilon_{m^\prime} \tau}
             P({\bf k}-{\bf k^{\prime}}, \epsilon_{m} -\epsilon_{m^{\prime}})
\point \nonumber \\
\end{eqnarray}

The linearized Eliashberg equation,
  which gives the onset of the finite value of 
  $\Sigma^a ({\bf k}, \epsilon_m )$,
  is expressed as
 \cite{Eliashberg} 
\begin{eqnarray}
  \label{eq:Eliashberg}
& & \lambda \Sigma^a ({\bf k}, \epsilon_m )
    =\sum_{{\bf k}^\prime , \epsilon_{m^\prime}} K({\bf k}, \epsilon_m ;{\bf k}^\prime , \epsilon_{m^\prime}) 
    \Sigma^a ({\bf k}^\prime , \epsilon_{m^\prime}) \virg
                \\
& &K({\bf k}, \epsilon_m ;{\bf k}^\prime , \epsilon_{m^\prime})
  =-\frac{T}{N_L} 
   P({\bf k}-{\bf k}^\prime , \epsilon_m - \epsilon_{m^\prime}) 
    \vert  G^0 ({\bf k}^\prime , \epsilon_{m^\prime}) \vert ^2 
     \nonumber
      \point
\end{eqnarray}

The SC transition temperature $T_{\rm c}$ 
is given by the condition, $\lambda=1$.
Since the characteristic momentum dependence of 
$P({\bf q}, \omega_l )$   is essentially  independent of $\omega_l$, 
and $P({\bf q}, \omega_l )$ decreases monotonously 
with $\vert \omega_l \vert$,
 eq. (\ref{eq:Eliashberg}) is calculated 
  by assuming a form given by  
\begin{eqnarray}  
 \label{eq:self_approx}
\Sigma^a(\bk,\epsilon_m)
 =
   \Sigma^a_1(\epsilon_m)
   \Sigma^a_2(\bk) 
   \virg  
\end{eqnarray}  
where $\Sigma^a_1(\epsilon_m)$
 is evaluated by  a sample  average 
   with some choices of  $\bk$.   
We solve the linearized Eliashberg equation numerically 
 by using an iterative method 
  where the symmetry in ${\bf k}$-space 
 is  rigorously determined 
  without the assumption for the  pairing symmetry.

 We examine the role of the spin and charge fluctuations  
  by varying $U (\geq 0)$ and $V (\ge 0)$. 
From eqs. (\ref{eq:chi_s}) and (\ref{eq:chi_c}),
  the spin and charge susceptibilities are calculated. 
For $U \not=0$  and  $V=0$, 
 the spin fluctuation is much larger than the charge fluctuation
 as seen from the solid line in Fig. 1. 
 This is because   the magnetic state is expected 
  by the repulsive on site interaction   
  leading to  the enhancement of 
   the spin fluctuation, $\chi^s$.  
   The nesting condition at 3/4-filling  gives rise to  $\chi^s$
   being  dominant   
   in the interval region of $(q_x,q_y)= (\pi,0)$ and $(\pi,\pm \pi)$, 
   while the peak of $\chi^s$ at 1/2-filling 
   appears at $(\pi,\pi)$.  
 For $V \gg U$, the charge susceptibility, 
  $\chi^c$,  becomes much larger than $\chi^s$,
    since   $V$ induces CDW 
   (or charge ordering) and 
     then the enhancement of the charge 
    fluctuation.
In this case, as shown by the dashed line in Fig. 1, 
   the peak of $\chi^c$ appears as the ridge connecting two points 
   of   $(\pi, \pi - b)$ and  $(\pi - b , \pi)$  with  $b \sim \pi /2$
 in a quarter  plane of $(q_x,q_y)$.
   This comes from the effect of  $V(\bf q)$ in eq.(\ref{eq:chi_c})
    exhibiting a peak at $(\pi,\pi)$ 
in addition to that of  the nesting condition.

\begin{figure}[tb]
\begin{center}
\vspace{2mm}
\leavevmode
\epsfysize=7.5cm\epsfbox{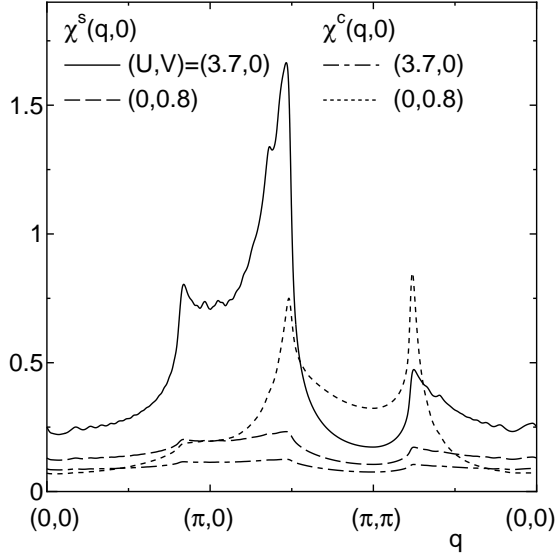}
 \vspace{-3mm}
\caption[]{
Spin ($\chi^s$) and charge ($\chi^c$) susceptibilities 
for $(U,V) = (3.7,0), (0,0.8)$ where $T=0.01$.
  }
\end{center}
\end{figure} 
   Based on Fig. 1, we calculate the pairing interaction, 
   $P(\bq, \omega_l)$, which is given by eq.(\ref{eq:pairing}). 
   For $U \not=0$ and  $V=0$, 
    $P(\bq,0)$ is always positive and 
   shows the large magnitude around $(\pi ,0)$ 
     with the peak located 
     between $(\pi,0)$ and $(\pi,\pi)$ 
     as seen from the solid line of Fig. 2. 
     The $\bq$ dependence of $P(\bq,0)$ is similar to 
      that of $\chi^s (\bq,0)$ in Fig. 1 since 
      eq.(\ref{eq:pairing}) leads to 
      $P(\bq,0) \propto \chi^s (\bq,0)$ for $V=0$
       and large $\chi^s$. 
   However, for  $V \not= 0$,  $P(\bf q)$ becomes  different from 
  charge and/or spin susceptibilities 
   since 
    the second line of 
    eq.(\ref{eq:pairing}) leads to $P(\bq)$ having 
     the sign opposite to $\chi^c$. 
  With increasing $V/U$, 
  $P(\bq)$ begins to decrease 
   especially in the region of $(\pi,0)$ and $(\pi,\pi)$ 
 where the peak of  $P(\bq)$ is reduced noticeably  
and the dip around $(\pi,\pi)$ grows leading to the attractive interaction.  
   Thus it turns out that, with increasing $V$,   the main part 
   of pairing interaction 
      becomes attractive and takes a broad dip 
       around $(\pi,\pi)$ but  
   still remains repulsive 
    in the region being away from $(\pi,\pi)$.      
 Apparently, such an attractive interaction indicates the s-wave pairing 
 but our obtained result differs from such a pairing  
  as shown later.

\begin{figure}[tb]
\begin{center}
\vspace{2mm}
\leavevmode
\epsfysize=7.5cm\epsfbox{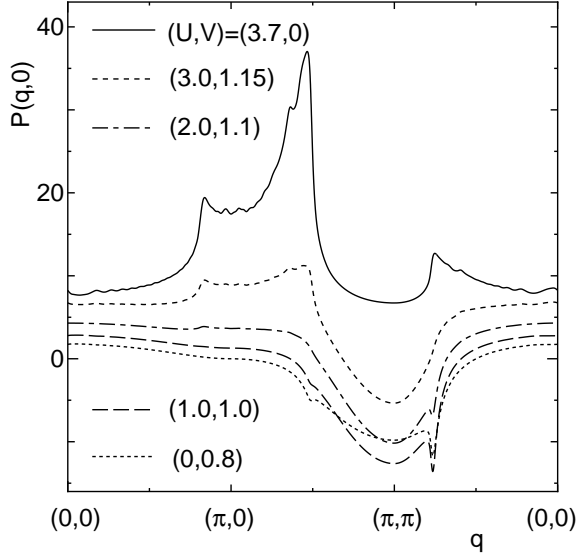}
 \vspace{-3mm}
\caption[]{
Pairing interaction $P(\bq,0)$ at $T=0.01$ 
 for $(U,V) =$ (3.7,0) (solid curve), (3.0,1.15) (short-dashed curve), (2.0,1.1) (dot-dashed curve), (1.0,1.0) (dashed curve)
 and (0,0.8) (dotted curve).
  }
\end{center}
\end{figure} 
The onset  for  the SC state 
  is obtained by calculating eq.(\ref{eq:Eliashberg})
   numerically.  
   The parameters $U$ and $V$ are calculated  so that 
    the maximum eigen value of $\lambda$ 
    may  become  unity 
     with the fixed $T$.  
    In Fig. 3, 
    the solid curve corresponding to the onset of the SC state
     is calculated for  $T = 0.01$. 
 On the  dashed lines,  
   either $\chi^s (\bq,0)$ or $\chi^c (\bq,0)$ diverges
    at $T=0.01$,
    i.e., 
   either the incommensurate SDW 
     or the incommensurate  CDW
    is obtained for $(U,V)$ in the right hand side 
    or the upper side of the respective line.   
  The region enclosed by the solid line, the horizontal axis  and 
   the vertical axis corresponds to the normal state. 
  The SC state 
   is obtained 
   in  the region enclosed by solid line and dashed two lines.
  The critical value of $V$ for the onset of the SC state 
     increases with increasing $U (\lsim 3)$. 
   This comes from the competition between 
   the spin fluctuation and the charge fluctuation 
   where the sign of  the respective pairing interaction 
    is  opposite each other as shown in Fig. 2. 
  Thus it is found that 
   the pairing interaction corresponding to 
     the onset of the SC state 
   is mainly determined by the charge fluctuation 
   for $U \lsim 3.5 $ 
    while the spin fluctuation is dominant for $3.5 \lsim U \lsim 3.7$, 
     i.e.,  close to the upper bound of $U$.    
   The SC region is enlarged 
    with decreasing temperatures since 
     the solid curve moves toward  the origin (0,0) 
      rapidly compared with that of the dashed line.

\begin{figure}[tb]
\begin{center}
\vspace{2mm}
\leavevmode
\epsfysize=7.5cm\epsfbox{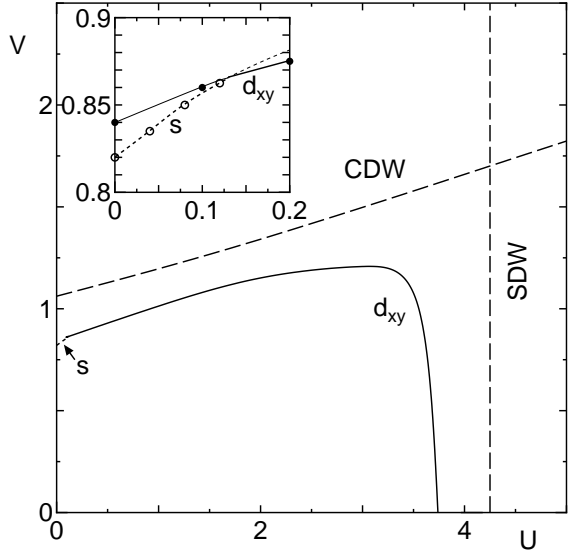}
 \vspace{-3mm}
\caption[]{
Phase diagram on the plane of $U$ and $V$ at $T=0.01$ 
where the superconducting state (SC) is obtained 
 in the region between the solid curve and the dashed curves. 
 The solid curve denotes the onset for the $d$-wave ($d_{xy}$) SC state 
 and the dotted line denotes that for the s-wave SC state. 
  In the inset,  the detail close to $U=0$ is shown 
  where solutions for both $s$ (dotted line) and $d_{xy}$ 
   (solid  line) exist for  $U < 0.2$.   
  }
\end{center}
\end{figure} 
In order to examine the symmetry of the order parameter 
 of  the SC state,  
 $\bk$ dependence of the quantity $\Sigma^a(\bk,\i \pi T )$,
is shown in Fig. 4 for three kinds of parameters.
The solid line, dashed line and dotted line correspond to  
 $(U,V)$ = $(3.7,0)$, $(1.0,1.0)$ and $(0,0.8)$, respectively.
In the case of $(U,V)$ = $(3.7,0)$ and $(1.0,1.0)$, we found following 
two features : 
 (1) the sign of $\Sigma^a(\bk,\i \pi T )$ in the region of $k_y >0$ are different from that of $k_y <0$, (2) $\Sigma^a(\bk,\i \pi T )=0$ 
 on $k_x$   and $k_y$   axis and on the boundary of the Brillouin zone, 
 indicating the relation, 
 $\Sigma^a (k_x, k_y, \epsilon_m) 
   = - \Sigma^a(k_x,-  k_y, \epsilon_m) $
 and  
 $\Sigma^a(k_x, k_y, \epsilon_m) 
   = - \Sigma^a(- k_x,  k_y, \epsilon_m) $. 
   By noting the symmetry of the square lattice, 
 this state corresponds to  the $d_{xy}$-like symmetry. 
In the case of $(0,0.8)$, on the other hand, the sign of 
 $\Sigma^a(\bk,\i \pi T )$ remains  unchanged  
  except for the region  near $(\pi ,\pi )$ and 
 $\Sigma^a(\bk,\i \pi T )$ has  a maximum near $(0,0)$. 
 Thus the dotted line of Fig. 4 indicates 
   the $s$-like symmetry. In Fig. 3, such a $s$ state is shown by 
    the dotted line. 
 The $d_{xy}$ state in the case of $(U,V)$ = $(3.7,0)$ is reasonable 
  due to the repulsive interaction 
    located around $\bq =(\pi,0)$
 and   seems to be consistent with that of 
   Kondo\cite{Kondo},
    who 
    treated the $U$-term perturbationally. 
 For $V$ being large 
  with the moderate strength of $U$ (dashed curve of Fig. 4), 
  the SC state is still given by $d_{xy}$
 although   the paring interaction of Fig. 2 is almost  attractive. 
   This is understood as follows. 
 Both $d_{xy}$ state and $s$ state can gain the the energy 
   from the attractive interaction close to $(\pi,\pi)$ 
    although the latter is slightly larger than the former 
     due to the absence of the node. 
   However $d_{xy}$ state can be  chosen due to the energy gain from the 
   the repulsive interaction around $(\pi,0)$, which remains  
    for the case of the dashed curve.     
   In the case of $(U,V)$ = $(0,0.8)$, Fig. 2 shows     
  that the repulsive interaction 
  around $(\pi,0)$ becomes sufficiently small 
  leading to $s$ state.. 
 The inset of Fig. 3 shows that  
  $d_{xy}$ state still exists as a metastable state at $U=0$. 
It is expected that  the difference of the Helmholtz free energy 
 between $s$ state and $d_{xy}$ 
 state is small for the case of small $U$.    
We note that there is no node for both $d_{xy}$ state and 
 $s$ state on the Fermi surface  
 of the present case of the 3/4-filled band. 

Here,  we analyze $\Sigma^a(\bk,\i \pi T )$  by using the Fourier
 transformation given by 
\begin{equation}
\Sigma^a(\bk,\i \pi T ) =\sum_{mn}c_{mn}\phi_{mn}({\bf k}) \virg
\end{equation}
with 
$
\phi_{mn}({\bf k})={\it e}^{{\it i}m k_x a} {\it e}^{{\it i}n k_y a}
/(2\pi)
$. 
In the case of $(U,V)$ = $(3.7,0)$, the ratio of the $d_{xy}$-component 
 (i.e., $\phi_{d_{xy}} ({\bf k}) 
 = 2 \sin k_x a \sin k_y a$) 
 to other components with larger momentum is given by 
  $1 : 0.50$, while  $s$- and  
 $d_{x^2 -y^2}$-component are absent.
In the case of $(U,V)$ = $(1.0,1.0)$, the corresponding 
 ratio is $1 : 0.37$.
For case of $ U \simeq 0$, 
 such a ratio remains approximately the same, and 
the $s$- and   $d_{x^2 -y^2}$-components also do not exist, 
 whereas the $s$-state is located very close to that 
  of the the $d_{xy}$-state   in the $U$-$V$ plane.
For the case of $(U,V)$ = $(0,0.8)$, the extended $s$ state
 shows  three components consisting of the $s$-component 
($\phi_s ({\bf k}) =1$), an extended $s$-component 
($\phi_{{\rm ext}-s} ({\bf k}) =\cos k_x a + \cos k_y a$) and
other components with larger momentum, 
 where the respective ratio   is given by $1 : 0.59 : 0.39$.
We note that 
$\Sigma^a(\bk,\i \pi T )$ becomes negative  near $(\pi ,\pi )$.

\begin{figure}[tb]
\begin{center}
\vspace{2mm}
\leavevmode
\epsfysize=7.5cm\epsfbox{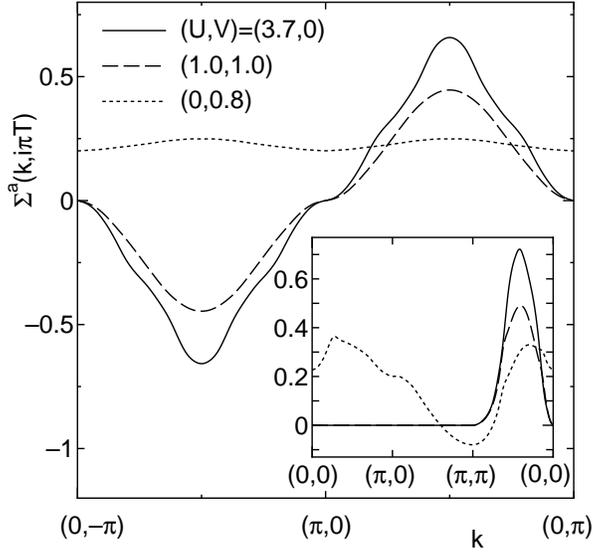}
 \vspace{-3mm}
\caption[]{
Self-energy for the SC state , $\Sigma(k, \i \pi T)$
  for $(U,V) =$  (3.7,0)(solid line), 
   (1.0,1.0) (dashed line) and (0,0.8) (dotted line) where $T=0.01$.
 The inset shows  the behavior in the quarter plane indicating that 
  the sharp peak  around $(\pi/2,\pi/2)$ for the $d_{xy}$ state  
   and the broad  peak around (0,0) for $s$ state.
  }
\end{center}
\end{figure} 
Finally we examine the SC by varying temperature in Fig. 5.
The frequency dependence in eq.(\ref{eq:self_approx})
  is discarded since such an effect becomes small with increasing temperature. 
  With increasing $T$ 
  the critical values  of $V$ for onset of both the SC state and the CDW state 
  increase  at low temperatures ($T \lsim 0.1$) and 
  at high temperatures ($0.6 \lsim T$), 
   while the value of $V$ decreases at the intermediate temperatures
    ($0.1 \lsim T \lsim 0.6$). 
 Such a reentrant transition of the CDW state and the SC state
   as the function of temperature,
   which resembles that of reference 8, 
 originates in the property of eq. (\ref{eq:chi_c}). 
 At low temperature, 
 the maximum of eq.(\ref{eq:chi_c}) is given by  
  the nesting wave vector of $\chi^0(\bq,\omega_l)$, 
 which is away from $(\pi,\pi)$
 as shown by the dotted curve of Fig. 1.  
 With increasing temperature, $\chi^0(\bq,\omega_l)$ as the function of 
  $\bq$ becomes broadened. 
   The nesting wave vector begins to move toward  $(\pi,\pi)$  
  at $T \simeq 0.08$ and also at the intermediate temperatures
  due to a competition of $\chi^0(\bq,\omega_l)$ and $|V(\bq)|$,
 which  results  in the increase of eq.(\ref{eq:chi_c}).
   At high temperatures,  eq.(\ref{eq:chi_c}) with 
    a maximum at $\bq= (\pi,\pi)$ decreases 
    due to the further  broadening of $\chi^0(\bq,\omega_l)$.

\begin{figure}[tb]
\begin{center}
\vspace{2mm}
\leavevmode
\epsfysize=5.7cm\epsfbox{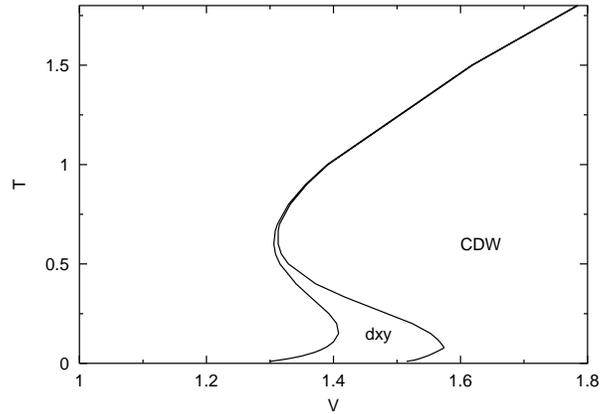}
 \vspace{-3mm}
\caption[]{
Phase diagram on the plane of $T$ and $V$,
where $U = 3$. 
  }
\end{center}
\end{figure} 
In summary,  
 we have examined the effect of nearest neighbor repulsive interaction 
 on the SC state for the 2D extended Hubbard model with 
a square lattice at 3/4-filling. 
We found the SC state with $d_{xy}$ symmetry 
  for both cases of  $U$  and   $V$, 
   although the pairing interactions 
 are  repulsive for $U$ and  attractive  for $V$. 
 The spin fluctuation induced by $U$ 
  gives the SC state with the symmetry being the the same as that
   of  the charge fluctuation induced by $V$.
However 
 the onset of the SC state needs  larger magnitudes 
  of interactions of  $U$ and $V$
 since these two fluctuations are incompatible as seen from Fig. 3.  
Thus it is interesting to extend  
 the present study to the case of  
  the SC state of organic conductors, in which 
   exotic states may be expected due to  
     the variety of  transfer energies. 
 
We are grateful for the financial support from a Grant-in-Aid for 
Scientific Research on Priority Areas of Molecular Conductors 
(No. 15073213 and 15073210) 
from the Ministry of Education, Science, Sports, and Culture, 
Japan
and 
for  Scientific  Research  from the Ministry of Education, 
Science, Sports and Culture, Japan (No. 14740208).


\end{document}